# Thermographic Laplacian-Pyramid Filtering to Enhance Delamination Detection in Concrete Structure

Chongsheng Cheng[1], Ri Na[2], and Zhigang Shen[3]

[1] Durham School of Architectural Engineering and Construction, University of Nebraska-Lincoln, 122 NH, Lincoln, NE 68588; e-mail: cheng.chongsheng@huskers.unl.edu

[2] Department of Civil and Environmental Engineering, University of Delaware, 342B Du Pont Hall, 127 The Green, Newark, DE 19716; e-mail: nari@udel.edu

[3] Durham School of Architectural Engineering and Construction, University of Nebraska-Lincoln, 113 NH, Lincoln, NE 68588; e-mail: shen@unl.edu

**Abstract:**

Despite decades of efforts using thermography to detect delamination in concrete decks, challenges still exist in removing environmental noise from thermal images. The performance of conventional temperature-contrast approaches can be significantly limited by environment-induced non-uniform temperature distribution across imaging spaces. Time-series based methodologies were found robust to spatial temperature non-uniformity but requires extended period to collect data. A new empirical image filtering method is introduced in this paper to enhance the delamination detection using blob detection method that originated from computer vison. The proposed method employs a Laplacian of Gaussian filter to achieve multi-scale detection of abnormal thermal patterns by delaminated areas. Results were compared with the state-of-the-art methods and benchmarked with time-series methods in the case of handling non-uniform heat distribution issue. Tor further evaluate the performance of the method numerical simulations using transient heat transfer models were used to generate the 'theoretical' noise-free thermal images for comparison. Significant performance improvement was found compared to the conventional methods in both indoor and outdoor tests. This methodology proved to be capable to detect multi-size delamination using single thermal image. It is robust to non-uniform temperature distribution. The limitations were discussed to refine the applicability of the proposed procedure.

## 1 Introduction

Thermography as a nondestructive evaluation (NDE) method has been used for decades to detect sub-surface delamination of concrete structure. The principle of detection is based on the difference in temperature gradient between intact and debonded area on the surface caused by different areal thermo-physical properties (Ibarra-Castanedo et al. 2017). As a result, the surface temperature for debonded area was expected higher than the intact area during the heating phase and lower during the cooling phase of a day (Hiasa et al. 2017). By differentiating the hotter or cooler regions from surroundings in the thermal image, the suspected area could be revealed and then engineering judgement could be made. Thus, the current practical field implementation focused on locating the potential defect area and estimating the approximate delamination range. Concerning estimating the depth of delamination using thermography, satisfactory results were found only under experimental condition with controlled artificial heating source (Arndt 2010; Dabous et al. 2017; Hiasa et al. 2017; Milovanović et al. 2017; Omar and Nehdi 2017). On the other hand, under the natural condition, the environmental variations including solar radiation intensity, clouding, shadowing and concrete surface heterogeneity (e.g. debris, texture, and color difference) negatively affects the detectability. In addition, non-uniform heat distribution across space was observed as a misleading factor for decision making when the threshold based method was used (Sultan and Washer 2017). Due to the two-dimensional nature of an image, the effect of temperature variation accounted by sub-surface and surface are superimposed and projected as a 2D representation which makes the direct interpretation become difficult in some cases. Thus, advanced image processing methodology for content analysis is needed to facilitate the thermal feature recognition for delamination detection.

Existing field methods found in the literature could be summarized into two major groups: the contrast-based methods and temporal variation-based methods. Contrast based methods use hard or soft threshold to distinguish debonded area from intact areas directly based on the spatial temperature difference. Due to its simplicity and only requiring a single thermal image, it was widely used in practical implementation (Dabous et al. 2017). This method heavily relied on personal judgement to select reference temperature of intact area and cutoff temperature contrast. Practically, the cutoff level of 0.5 °C was recommended by ASTM (D4788-03). Recently, further evaluation was conducted by Hiasa et al. (2017) in terms of different size and depth in mimicked defects using simulated model and recommended 0.4 °C in contrast as the lower bond for the certain detectability. On the other hand, Sultan and Washer (2017) deployed a reliability analysis to trade off threshold selection between false-positive and true-negative when non-uniform temperature distribution occurred across space. Base on their result, the optimum contrast of 0.6~0.8 °C was preferred. Alternatively, instead of selecting reference (intact area) and target (debonded area) temperature subjectively from experimental data, Hiasa et al. (2017) obtained the temperature contrast from a simulated model at corresponding time window and then calculated the target temperature by subtracting from captured thermal image. Also, Omar et al. (2017) implement the k-means clustering to group temperature features based on numerical density by predefining number of k centrodes practically to replace manual thresholding process.

Temporal variation-based methods focus on the temporal evolutional difference in temperature to distinguish debonded from intact areas. Pulse Phase Thermography (PPT) and Principle Component Thermography (PCT) were the two classical methods found efficient in both experimental and natural environment (Arndt 2010; Dumoulin et al. 2013; Ibarra-Castanedo et al. 2017; Milovanović et al. 2017). Since it primarily utilizes the information in the temporal dimension, it was referred as time series thermography in literatures. The key idea of the two methods was to project the temporal data from the time domain to a new domain so that the spatial variation could be re-aligned more properly associating to the thermo-physical properties of the material (more detailed review in section 2.1). Additionally, the reveal of the defect depth became more robust in terms of the issue for spatial non-uninform heat distribution.

Given the recent advancements of computer vision for image content analysis, and the absence of specific image processing method designated for delamination's thermal pattern identification, this study aims to develop an analytical procedure to breakdown the thermal image based on spatial band pass filtering to reveal and enhance delamination detection in both indoor experiment and outdoor natural conditions. In the following paragraphs a new empirical method is reported with enhanced detecting performance compared to existing methods.

## 2 Background

### 2.1 Time-Series Thermography for Concrete Sub-Surface Defect Detection

Time-series thermography was recognized as a robust methodology for subsurface defect detection implemented for concrete structures (Arndt 2010; Milovanović et al. 2017). Pulse phase thermography (PPT) is a Fourier transformation-based methodology used for sub-surface defect detection for a metallic material (Maldague et al. 2002). It was extended by Arndt (2010) to evaluate the delamination-like defects submerging in concrete structures under the indoor environment with artificial heating. Few efforts was found to explore the applicability under natural environment. Dumoulin and Averty (2012) and Dumoulin et al. (2013) revealed the inner structures (beam and girder underneath the deck) of a bridge with a calibrated system. Ibarra-Castanedo et al. (2017) used PPT for detecting subsurface features (e.g. hidden door and hidden window) for exterior walls. The mechanism behand was explained by Ibarra-Castanedo et al. (2017) that the daily solar loading behaved similarly to a modulated periodic heating in the laboratory and could be represented in a simplified form based on the thermal wave theory so that the periodic temperature data was then processable for above methods. Principle component thermography (PCT) is an alternative transformation-based method implemented by Milovanović et al. (2017) for delamination detection of the concrete slab under the lab condition. This method converted the data from spatial-temporal space into variation-in-orthonormal space so that the primary information along time axis was reordered in terms of variation and could be used to distinguish intact from debonded areas (Ibarra-Castanedo et al. 2009). However, both methods required a lengthy period for data collection (such as couple days) to obtain the appropriate result which limited the feasibility for the short-term inspection which often required a rapid data collection and non-interruption to the normal use of the structure (such as in-serve bridge or road).

2.2  Laplacian of Gaussian (LoG) Filter for Blob Detection

Blob detection in the field of computer vision was referred as detecting brighter or darker regions from surroundings. One of the common detectors named as Laplacian of Gaussian (LoG) operator was found to be effective for detecting blobs within an image at a specific scale and it was generally applied for medical image processing for cell identification (Zhang et al. 2015). The key process of LoG was to define the optimal scale so that a specific size of blob could return a strong response (positive value). The scale which was also referred as the standard deviation (σ) of the Gaussian kernel was found a strong relationship with blob size (Kong et al. 2013). Here, defines the radius of the blob as $s$ and the standard deviation σ of LoG was then calculated as $\sigma = (s-1)/3$. Thus, this operator would perform well when the blob size was known and when all blobs were the similar size in the image. To detect different sizes of blobs, a multiscale approach was introduced based on scale-space theory (Lindeberg 1998) in the later subsection 2.3. On the other hand, a conventional form of LoG was symmetric so that it worked well for circular blobs but performed unsatisfactorily when the blob had irregular shapes. A generalized LoG filter was developed by Kong et al. (2013) to improve the detection performance in terms of different sizes and shapes and the formulation was given below (Eq. 5):

$$\nabla^2 G(x,y) = \frac{\partial^2 G}{\partial x^2} + \frac{\partial^2 G}{\partial y^2} \quad (1)$$

$$\frac{\partial^2 G}{\partial x^2} = A[(2ax+2by)^2 - 2a] \cdot e^{-(ax^2+2bxy+cy^2)}$$

$$\frac{\partial^2 G}{\partial y^2} = A[(2bx+2cy)^2 - 2c] \cdot e^{-(ax^2+2bxy+cy^2)} \quad (2)$$

Where $x$ and $y$ denote the 2D space coordination, $A$ is the normalization factor, $a$, $b$, and $c$ are the coefficients controlling the shape and orientation.

$$a = \frac{\cos^2\theta}{2\sigma_x^2} + \frac{\sin^2\theta}{2\sigma_y^2}$$

$$b = -\frac{\sin 2\theta}{4\sigma_x^2} + \frac{\sin 2\theta}{4\sigma_y^2}$$

$$c = \frac{\sin^2\theta}{2\sigma_x^2} + \frac{\cos^2\theta}{2\sigma_y^2} \quad (3)$$

The normalized factor here was denoted as $A = (1+\log(\sigma_x)^\alpha) \cdot (1+\log(\sigma_y)^\alpha)$ by Kong et al. (2013) that **α** here is the positive number to control the blob-center eccentricities. After having the LoG filter generated, convolve the filter with desire image would achieve the blob detection through tuning the hyperparameters $(\sigma_x, \sigma_y, \theta)$ to get the correct scale and orientation. The issue remained was then for the multi-scale detection.

2.3 Image Pyramid and Scale Space Representation

Image pyramid is a common tool for digital image process to handle the multi-scale feature presentation in an image. It is based on the observation that objects represented in an image are meaningful when they are at specific scales (Lindeberg 1994). This scale defines the detail level of the object represented in the image space and it reveals the fact that the feature of the object represented in the image is scale dependent at the given space and it forms the formal scale space theory (Lindeberg 1994). When it comes to the machine for automatic feature detection, the filter (such as LoG for blob detection) needs to change its own scale to fit the different scales of object in the image. Thus, we either need to variate the scale of the filter for the feature in the image or adjust the scale of the feature in the image to fit the size of the filter. The image pyramid then was proposed as one of the solutions that could achieve the purpose. It changed image scale through expanding and reducing by the power of two after convolving with a fixed gaussian kernel (Adelson et al. 1984). As the result, the processed image would reveal the sub-banded behavior as the pyramid structures (Heeger and Bergen 1995). The advantage of this process was computational efficient due to exponentially decrease of the image size and the correspondence to different spatial-frequency bands (Lindeberg 1994).

With the development of scale space theory (Lindeberg 1994), the continuous feature representation was theoretically proofed to be achieved by using different scales of gaussian kernels and their derivatives. This supported that the varied sizes of Gaussian filters could be used as holo-scale filtering process for a fixed-size image. However, we may face the issues that the actual blob size is unknown beforehand, and blobs may randomly place in the image so that multiple estimations of $\sigma$ are needed for detection in different sizes. To address above short come, an empirical process is proposed in the following section.

## 3 Proposed Methodology

The main idea of Laplacian of Gaussian filter is based on the observed phenomenon that the thermal pattern of delamination can be treated as hot or cool regions which is similar to the blob features in the image. They both represent the island-liked regions with different shapes and sizes. Additionally, the thermal image of the concrete structure in the field often faces the issue of non-uniform heating and surface texture disturbing so that the conventional threshold-based method suffers uncertainties for supporting the engineering judgement. The essence of using Laplacian of Gaussian filter is based on the positive response occurred at the local extrema when the size of blob is close to the detector size. In the paper this property is used to enhance the detection performance from the background in the thermal image.

The proposed method will be divided into two parts: 1) generation of LoG filter, and 2) convolution with pyramid structures. There are three controlling parameters that determines the generation of LoG filter: $\sigma_x, \sigma_y, \theta$. Figure 1 illustrates the generalized LoG filter with different scales and orientations. Those variations had been found related to blob shape and orientation in Kong et al. (2013). Second part is to down sample image by pyramid process to provide different scale information. Lastly, convolve the selected LoG filter with pyramid images to get desired detection at different scale levels.

The steps are as follows (Figure 2):
(1) Generate the LoG filter $h(\sigma_x, \sigma_y, \theta)$ by defining sigma and orientation;
(2) For *i* equals desired levels *L*, *do* pyramid "reduce" process for raw image $I$ by $2^i$ times to have $I_i$;
(3) *Then* convolving reduced image $I_i$ with $h(\sigma_x, \sigma_y, \theta)$ to have $I_i'$;
(4) *Do* pyramid "expand" process for $I_i'$ by $2^i$ times to have $I'$.

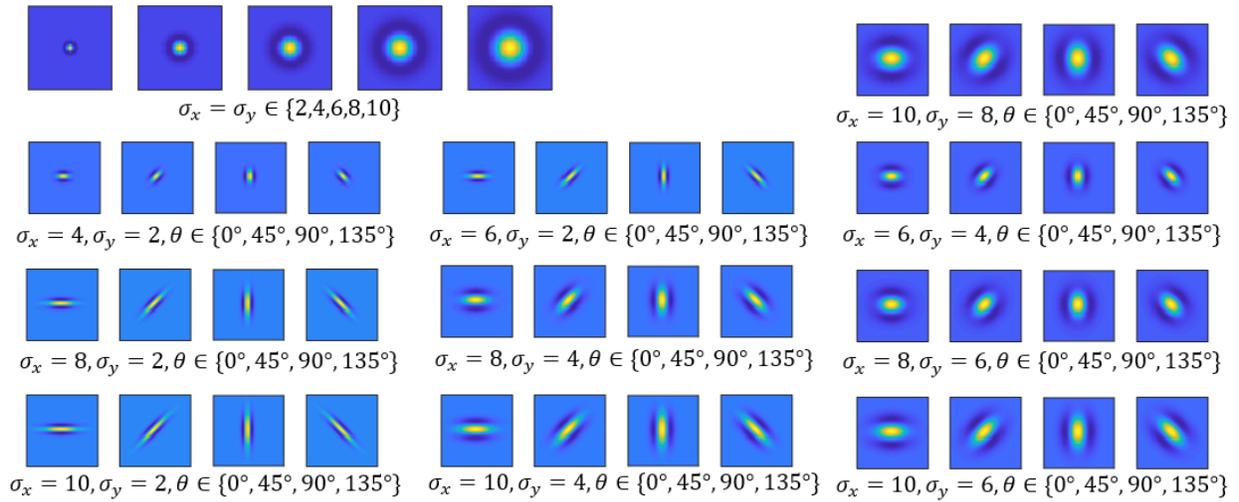

Figure 1. generalized Laplace of Gaussian filter with different scales and orientations

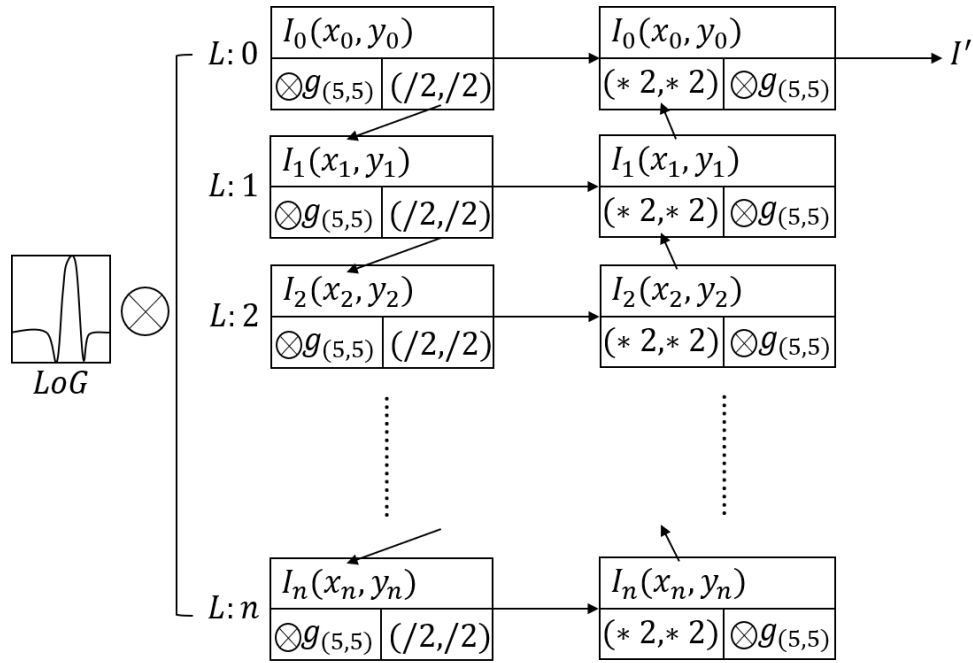

Figure 2. Architecture of the proposed process

## 4 Experiments

### 4.1 Experimental Design

A concrete slab was designed and constructed to mock a typical concrete bridge deck. The design followed the empirical deck design procedure from *Bridge Office Policies and Procedure* (NDOR) but with single layer of reinforcement. Figure 3 (a) shows the layout of reinforcement. To mimic the delamination, the Styrofoam was casted into varied sizes and located in different depths (Figure 3 (b)). Figure 4 (a) shows the as-built configuration of the Styrofoam in the formwork. The thickness of foam is 5/32 inch with thermal conductivity of 0.03 W/(m·K) to simulate the delamination in real bridge deck which is identified as a thin layer of air with thermal conductivity around 0.02 W/(m·K). Figure 4 (b) shows the finished surface of the slab after 28 days' curing and the heterogenous surface textures are observed.

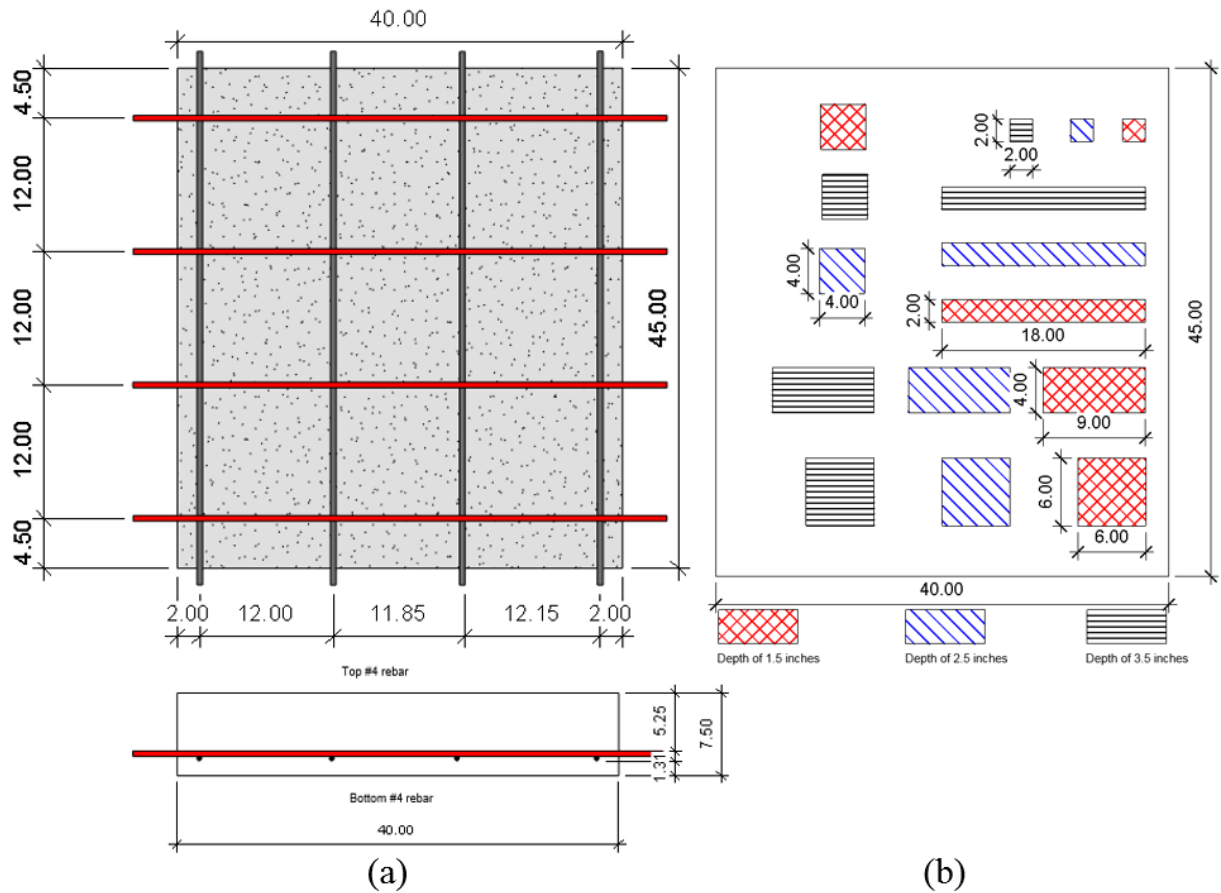

Figure 3. (a) Concrete slab and reinforcement layout; (b) mimicked delamination layout

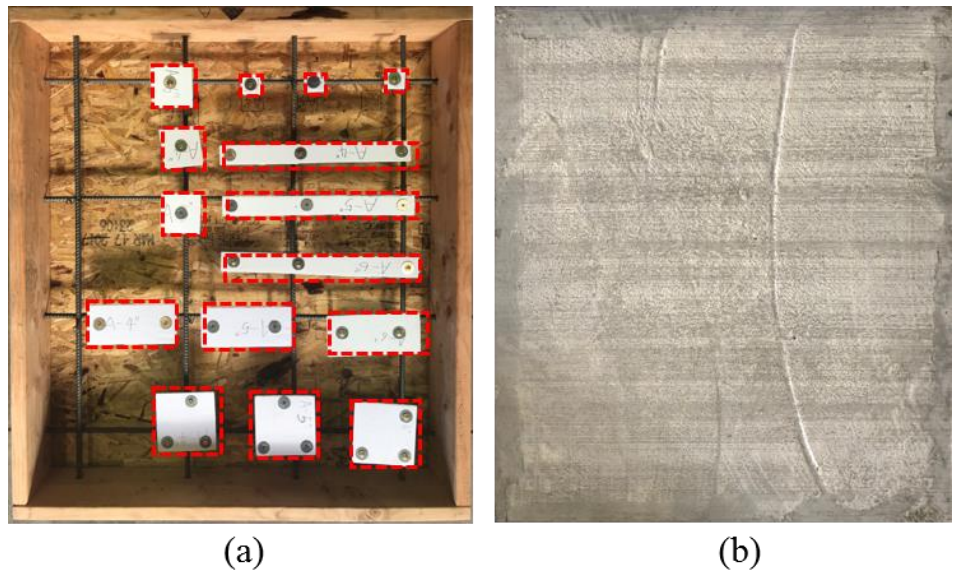

Figure 4. (a) as-built of mimicked delamination; (b) finished surface of concrete slab

4.2 Experiment and Data Collection

The experiment was conducted in both indoor and outdoor environment to illustrate the applicability of the proposed method. Figure 5 shows the two heating sources: (a) halogen lamp; and (b) solar radiation in natural environment. In the lab study halogen lamps were used to simulate solar radiation in the indoor environment. The stable indoor temperature creates an ideal environment for baseline (with absence of many outdoor noises) observation for comparison with the observations in natural outdoor environment.

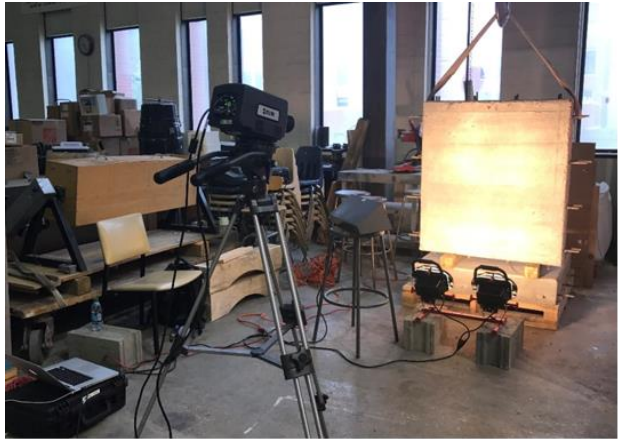 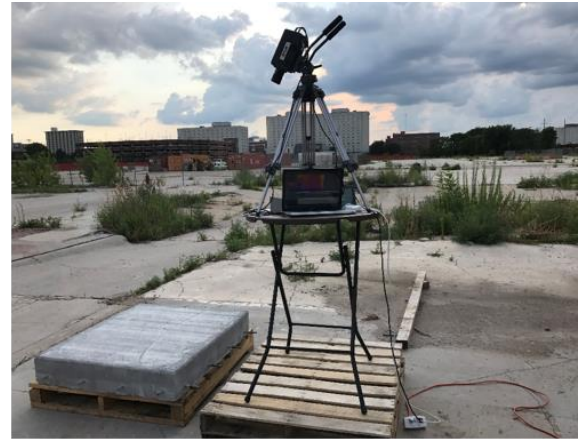

|(a)|(b)|

Figure 5. (a) indoor settings with halogen lamp as a heating source; (b) outdoor natural settings with solar radiation

The data were collected at the 0.2Hz sampling rate for the indoor and outdoor environment. Figure 5 (a) shows the slab was heated by halogen lamp for 220 minutes and then set for cooling for 139 minutes on July 14th, 2017. Both phases were recorded by the thermal camera (FLIR A8300). Figure 5 (b) shows the data collection outdoor on August 1st, 2017. The heating power and air temperature were controlled in indoor environment to keep consistent heating and stable boundary condition. Thus, the heat transient behavior was more stable (ideal) compared to the natural environment which the power of solar radiation and ambient temperature were both varying during a day.

The raw thermal image from above scenarios present the considerably low contrast for visual inspection in Figure 6. In the Figure 6 (a) and (b), it is shown the images recorded at 218-minute at the heating phase and 30-minute at cooling phase correspondingly for the lamp heating. Figure 6 (c) (d) and (e) show the images taken at 10:00am, 3:00pm, and 8:00pm under natural environment.

The observed raw images above exhibited variety of situations for the non-uniformed heat distribution that degraded the detectability due to the high background variation in temperatures. This effect comes from two major sources: (1) the trend of global temperature distribution is non-uniformed; (2) the scattered reflection received by the camera from the non-uniform surface texture. In the indoor condition, the causation is mainly due to the non-uniform of the heating source (Figure 6 (a) and (b)). Also, a strong reflection effect was observed in the heating phase by the lamp heating (Figure 6 (a)) which is due to the variation of surface emissivity (Vollmer and Möllmann 2010). In the outdoor environment, it is considered as the combination of above two effects. The non-uniform heating is mainly caused by different angles of sun during a day that heats the sample slab unevenly and the scattered reflection always existed due to surface textural variation.

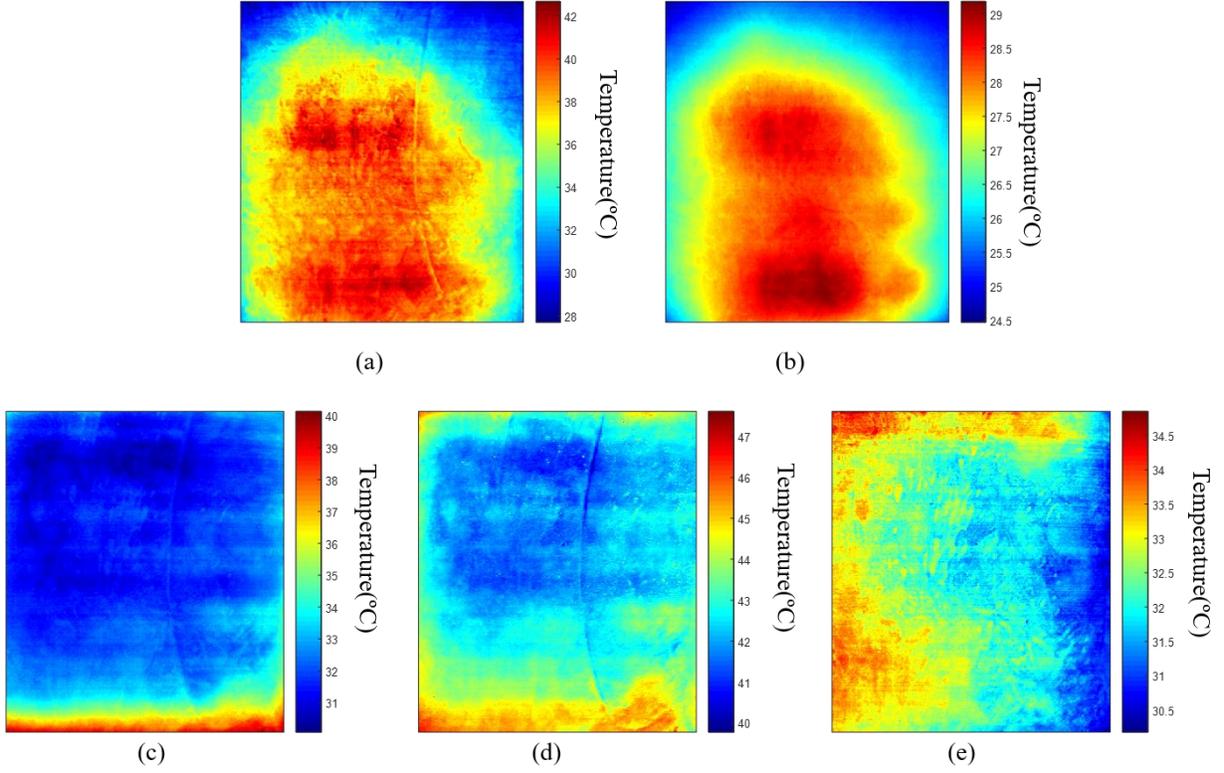

Figure 6. raw thermal images in three scenarios: (a) at heating phase in Figure 5 (a); (b) at cooling phase in Figure 5 (a); (c) (d) and (e) at 10 am, 3 pm, and 8 pm under natural environment in Figure 5 (b).

## 5   Result and Performance Comparison

### 5.1   Performance on Raw Image and Soft Thresholding based on Rectified Linear Unit

As shown in Figure 7, the performance of proposed procedure is significantly improved compared to the raw image from Figure 6 (b). With selecting the sigma size of 2 (pixel units), the LoG filter served as the detector to give positive response from the fine scaled image (L0) to the coarse scaled image (L4) when the target size was round of 14 (pixel units) in each level. In the Table 1, the size of pre-defined delamination was decreased correspondingly. For level 0 and level 1, the detector was reacting for the pattern that have a diameter of 0.924 inch and 1.848 inches theoretically (Kong et al. 2013). However, it was found there was a tolerance for the size sensitivity that approximated from *s/2* to *2s* (*s* refers the size parameter in section 2.2) which made the detector give a similar response for the pattern having the size from 7 to 28 roughly. This behavior was also observed in the level 2 to level 4 images. By summation of the processed images (from level 1 to level 4) would give integrated detection for multi-scale targets into a single image (Figure 7 L(1+2+3+4)). Since we are looking for the hot spot in the cold background (brighter blob in dark background), the summed positive response revealed in the image is the desired outcome. Thus, we adopt the concept of rectified linear unit from Glorot et al. (2011) as the activation function for our thresholding procedure. The rectifier function was defined as:

$$f(x) = \max(0, x) \qquad (4)$$

defines $f(x) = 0, when\ x \leq 0;\ f(x) = x, when\ x > 0$. By processing this function through all pixels in the image space, it will return the 0 for the input signal below 0 and the same value of input when the input signal larger than 0. This behavior was chosen to be beneficial for our purpose that we wanted to avoid the patterns which had negative response after procedure and kept the information for the patterns which had the positive response (Figure 7 L(1+2+3+4)>0). In the image, the mimicked delamination in depth of 1.5 and 2.5 inches with size of 2 by 18, 4 by 4, 4 by 9, and 6 by 6 as well as the depth of 3.5 with size of 6 by 6 were visually recognizable (shown in red box).

Table 1. Styrofoam sizes and scale change in different levels

| Size representation in pixel (inch per pixel) | 2-inch Styrofoam (pixel size) | 4-inch Styrofoam (pixel size) | 6-inch Styrofoam (pixel size) |
|---|---|---|---|
| L0: 0.066 | 30 | 61 | 91 |
| L1: 0.13 | **15** | 30 | 46 |
| L2: 0.26 | 8 | **15** | 23 |
| L3: 0.52 | 4 | 7 | **11** |
| L4: 1.05 | 2 | 4 | 6 |

5.2  Performance Comparing with State-of-the-art Methods

The proposed method performed better than existing methods based on contrast or density distribution (Figure 7. bottom) in the indoor environment. The threshold method (Figure 7 threshold) allows to screen the debonded area based on the pre-defined the temperature (such as 32 °C here). However, it had to take cautions when non-uniform temperature distribution existed in the imaging space mentioned by Sultan and Washer (2017). The method proposed by Hiasa et al. (2017) required the selection of temperature for sound area from experimental measurement and calculated the temperature difference from simulation at the same time window. Then the reconstructed image was represented in a range of 0 to 1 so that a cut-off value could be applied to distinguish debonded area (Figure 7 Hiasa et al. (2017)). K-means clustering was a density-based method implemented by Omar et al. (2017) for assessing delamination in bridge deck into different severity levels through pre-defining the number of k centroids. All three existing methods suffered low performance in terms of the concentrate heating procedure which developed a radial shape of temperature distribution. Under this circumstance, the previous methods revealed a strong reliance on the assumption that a uniform heating procedure was required. However, this pre-condition is often not practical in either laboratory or natural environment. It also found the proposed method applicable for a wide range of time windows during the cooling phase (Appendix Figure I). Additional comparison for lamp heating case could be seen in appendix Figure II. In the lamp heating case, it failed to indicate the location of mimicked defects and so as other methods except for time-series method. It was due to the strong reflection effect from surface radiation provided by lamps during the heating so that a consistent pattern was observed from beginning to the end through all the time (appendix Figure III). Often, a high-power heating source causes saturation of the thermal image during heating and so forth the usage was not recommended for active thermography (Ibarra-Castanedo and Maldague 2004).

5.3  Benchmarking with Time-Series Thermography

The proposed method presented a compatible detectability to the time-series thermography. The Pulse Phase Thermography (PPT) utilizes the properties of phase delay for the temperature evolution across debonded area and intact area (Ibarra-Castanedo and Maldague 2015) so that the distinguishable image appears at very low frequency band for concrete structure (Maierhofer et al. 2006). The phase image at 0.24 mHz was then selected here to demonstrate the best detectability of PPT and it was visually distinguishable for defects in depth of 1.5 inches with size 6 by 6, 4 by 9, and 4 by 4 (Figure7. PPT: marked by red boxes). However, using PPT requires the appropriate sampling rate and a long time to develop the transient heat conduction behavior. For the Principle Component Thermography (PCT), the targets at depth of 1.5 inches were clearly detected in the second principle image (Figure 7. PCT2 red boxes). Additional detection for the defect with depth in 1.5, 2.5 and 3.5 inches were found in the third principle image (Figure 7.PCT3: red boxes). Even though the distinguish information could be found in the first few principle images (Milovanović et al. 2017), the manual screening and selection was required. The detectability for lamp case was similar in the cooling phase (appendix Figure III). In the lamp heating case, the defects with size of 6 by 6 and 4 by 9 at depth of 1.5 inches were only visible in PCT and PPT (appendix Figure II). Overall, the proposed method provided a similar detectability benchmarking to time-series methods in cooling cases. However, the time-series methods required a sequence of data for process while the proposed method only required single thermal image.

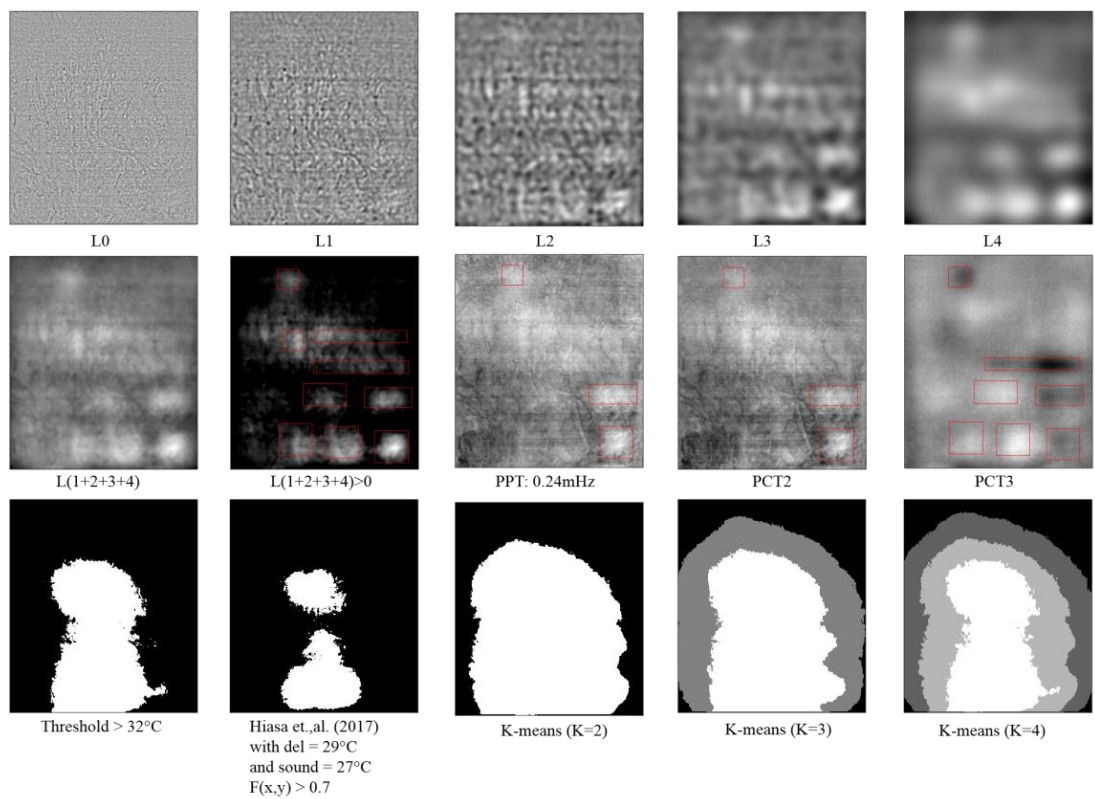

Figure 7. results of the indoor experiment

Figure 8 shows the processed images of the same slab under the natural environment. Three time-windows were selected to illustrate the best results in temperature inclining phase, peak phase, and declining phase (Figure 6 (c) (d) and (e) correspondingly). The authors examined all raw images for that day and found the location of pre-defined defect areas were not visually recognizable in all raw images. Existing methods suffered from the non-uniform heating where the east side of slab heat faster at morning, the south side during noon, and west side at evening. Thus, the performance is degraded and only the defect with size of 4 by 9 at depth of 1.5 inches were recognizable at the time window of 10 am and 3 pm (Figure 8 (b)(c) and (f)(g)). By processing with proposed procedure, more clean results were obtained. In Figure 8 (a), the debonded area could be recognized majorly for size of 4 by 9 and 4 by 4 at depth of 1.5 inches and partially for size of 4 by 9 and 4 by 4 at 2.5 inches depth and 2 by 18 at depth of 1.5 and 2.5 inches at 10 AM. A clearer recognition (Figure 8 (e)) for defects with size of 6 by 6 at depth 1.5 and 2.5 inches were observed for 3 PM. There was no feasible observation found at 8 PM due to unrecognizable thermal contrast (see Figure 10 in Section 6). Interestingly, different contrasts observed at different time windows and preferred time window for this case was afternoon at the summer. This finding is consistent with the temperature evolution curve developed by Hiasa (2016) which the maximum contrast could occur at afternoon. Lastly, it was unable to provide better results using time-series method primarily due to the random nature of the environmental factors. This finding convinced with the similar implementation for field detection in literature (Dumoulin and Averty 2012; Ibarra-Castanedo et al. 2017) which more than 3-days data was used.

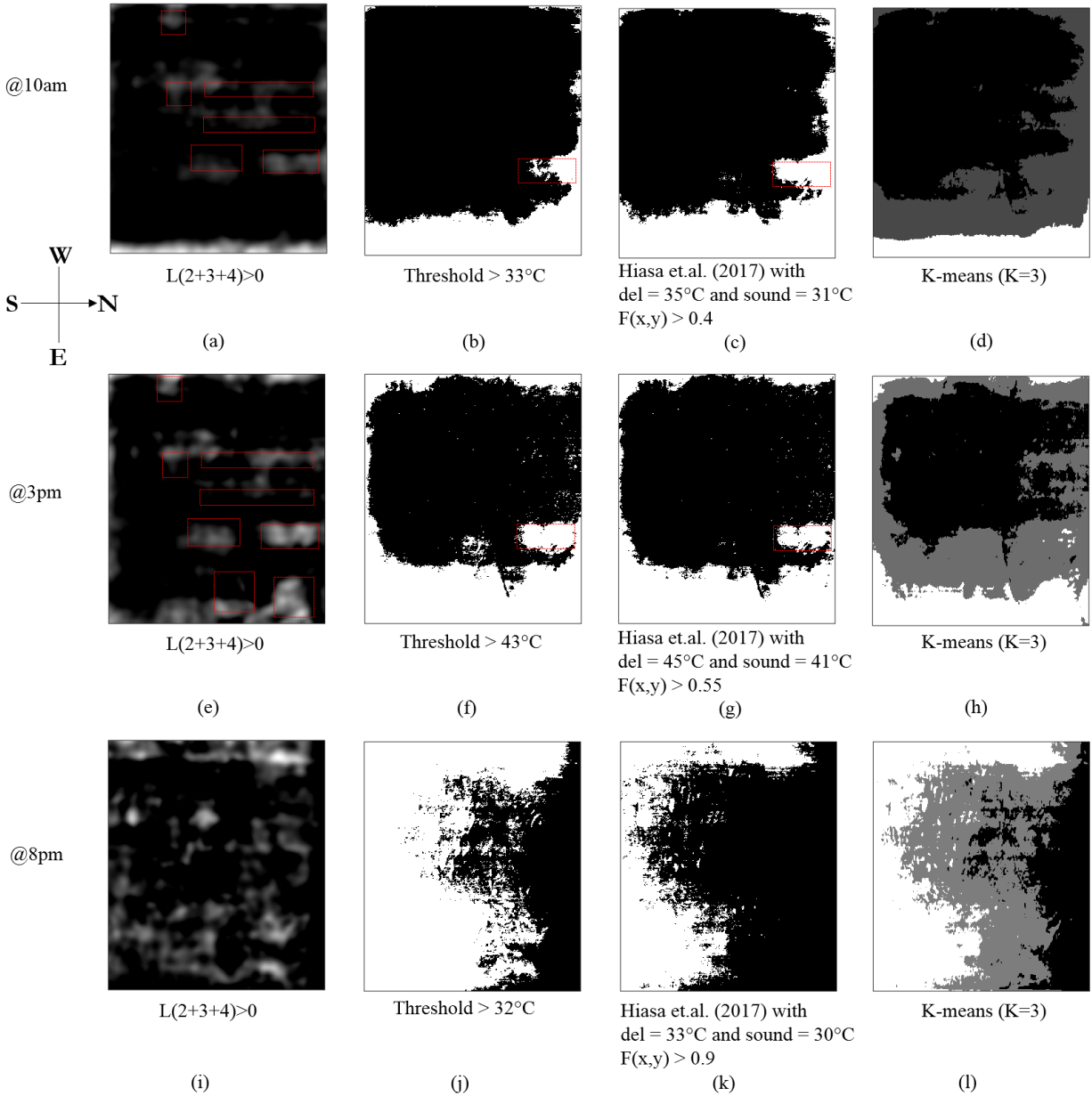

Figure 8. results from slab at outdoor

### 5.4 Interpretation for Processed Thermal Image

To further understand the outcome of processed image, a line profile (in the dash line location in Fig. 9) from the raw image and processed image were pulled out and compared in Figure 9. The dash lines represent the same position located in the raw images ((a) and (e)) and processed images ((d) and (h)) and were plotted in (b)(f) and (c)(g) correspondingly. Red dots in charts indicate the center location of buried Styrofoam. As shown in figure 9 (b), it is hard to use single threshold to distinguish all debonded areas based on temperature since the local hot regions are immersed in a larger non-uniform temperature trend caused by heating source. By processed with the proposed method, this effect was largely reduced so that more contrast was developed for local maxima from its surroundings (Figure 9(c)). Furthermore, a general threshold of 0 could be applied here to screening the hotter regions consistently. Thus, the more regulated visualization is presented Figure 9 (d). Comparing the raw image for sample outside at 3 pm (Figure 9(e)), the processed image considerably improved the contrast (Figure 9(g)). However, the performance was not as feasible as shown in indoor condition.

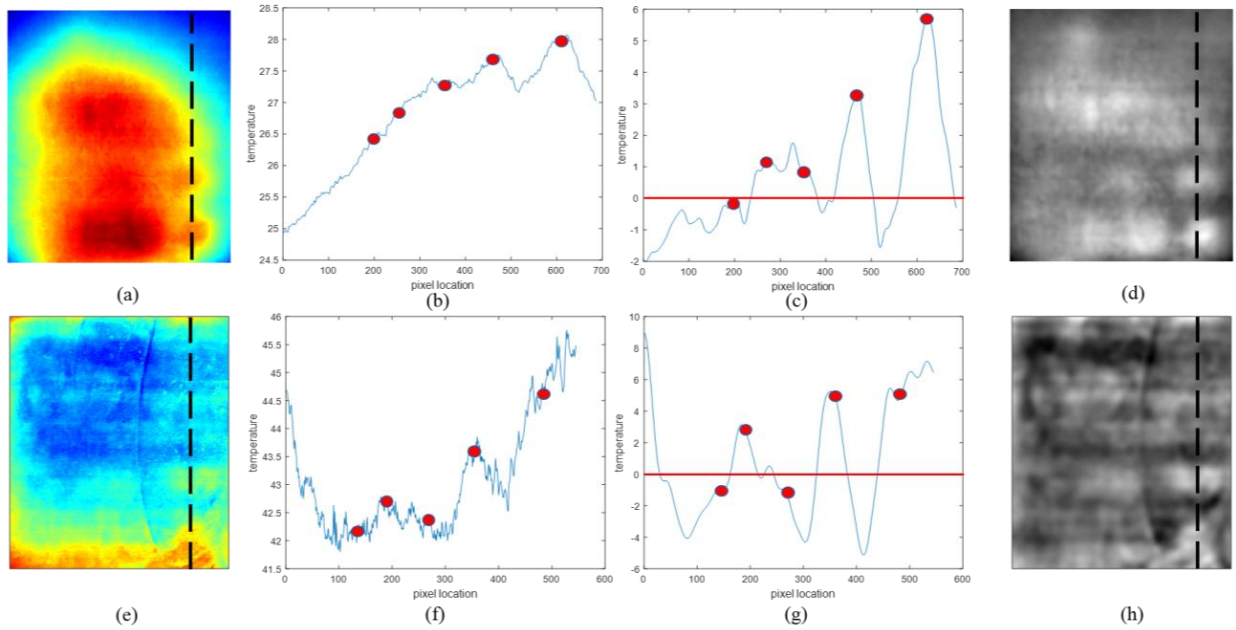

Figure 9. detailed comparison before and after processing; the red dots indicate the center location of mimicked delamination in the slab

## 6  Numerical Simulation Results for Further Baseline Comparisons

Numerical simulation has been validated for modeling the concrete slab with delamination in terms of its detectability by thermography under natural environment in recent researches. A series of studies conducted by Hiasa et al. (2017) demonstrated the usefulness for evaluating the detectability in terms of variate geometries of mimicked delamination and weather conditions. Thus, two numerical simulations (one indoor and one outdoor) were conducted to cross-check the detectability under indoor and outdoor environment. The model aims to simulate the transient heat transfers in both indoor and outdoor settings. The results are expected to provide a 'theoretical' noise-free benchmark for the surface temperature variations between the debonded and intact areas of the concrete slab. Since the uniformly distributed radiation assigned and homogeneous material surface with uniform emissivity, combing with smooth ambient temperature conditions, the calculated results could serve as the benchmark of maximum detectability for the experiment results.

In the simulation section, 3D transient numerical models of the indoor and outdoor experiments were created using Autodesk CFD 2017(Autodesk 2017). As an important Computational Fluid Dynamic (CFD) application, the radiation model works with the turbulence model such as K-epsilon model can support the simulation of mixed heat transfer including radiation, convection and conduction. It has been applied in several studies that involves solar radiation in the heat transfer process, such as analyzing the performance of solar energy collector (Gholamalizadeh and Kim 2014; Martinopoulos et al. 2010; Selmi et al. 2008). To simulate the radiative transfer process, the radiation model requires the transmissivity, and emissivity properties to be assigned to the material. All these physical parameters were assigned with the default material settings in the Autodesk CFD. The details of the heat transfer calculation method in the radiation model can be found in the Autodesk user's manual (Autodesk 2017). The geometry settings of the numerical simulations were exact same as the experiments, while the boundary conditions are listed in Table 2.

**Table 2**. The boundary condition of the solar heating model

| Parameter | Definition | Indoor Case Value | Outdoor Case Value | Unit |
|---|---|---|---|---|
| **Ambient air temperature, $T_a$** | The constant room temperature for the indoor case and hourly ambient air | 24 | NOAA data[1] (NOAA 2017) | °C or (°F) |
| | temperature for the outdoor case | | | |
| **Film Coefficient, h** | The heat transfer rate due to the convection | 20 | N/A² | W/m²/K |
| **Latitude $\varphi$ and longitude $\lambda$ of the locations** | Lincoln, NE | N/A | 40° 86' N, 96° 68' W | - |
| **Radiation Intensity, $\varphi_q$** | The radiation received on the concrete surface | 600 | N/A³ | w/m² |

**Note:**
1. Average air temperature during the last 5 minutes of the hour.
2. For the outdoor case, the computational domain includes both the concrete block and the air block. However, the air was only used for the simulation of the radiative heat transfer. The solar flux constant $E_e$ was lowered to 310 w/m² so as to provide the equivalent film coefficient due to the convection.
3. The solar radiation intensity is calculated by the solar ray tracing algorithm in Autodesk CFD based on the geographical location and time information.

For the indoor case, the ambient temperature $T_a$ was set close to the experimental setup for the lamp case. Due to the relative constant convection effect between slab surface and the ambient air, the simulation domain for the indoor case was only the concrete block (Figure 10a). The impact of natural convection was calculated by applying a constant film coefficient h. The radiation heat was simulated by assigning a consistent heat flux $\varphi_q$ in the first four hours.

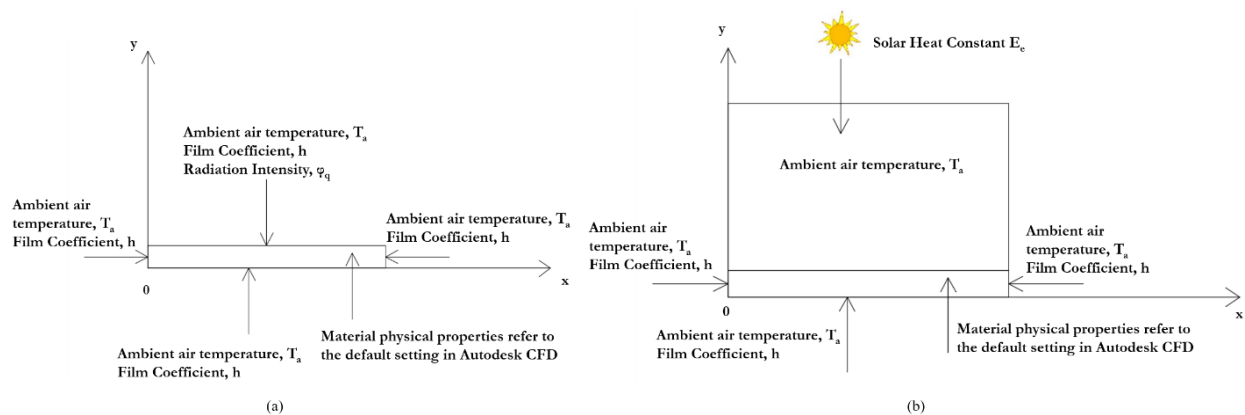

Figure 10. Schematic of computational domain (a) indoor case and (b) outdoor case

For the outdoor case, the simulation of outdoor experiment used a radiation model, so called solar heating model, which can include the radiation effect from the sun's ray. The computational domain contained both the concrete block and the surrounded air (Figure 10b). The ambient temperature $T_a$ applied in the simulation was from the weather data collected by the local weather station on August 1st, 2017 (Figure 11). The solar radiation intensity $\varphi_q$ is calculated by the solar ray tracing algorithm based on the given geographical location, time information, and the solar flux constant. Thus, the solar radiation $\varphi_q$ is not a fixed value. It varies with time and location and requires the time, the longitude $\lambda$ and latitude $\varphi$ of the location as input parameters. Please note that the computational domain of air was only used to simulate the radiative heat transfer in this case.

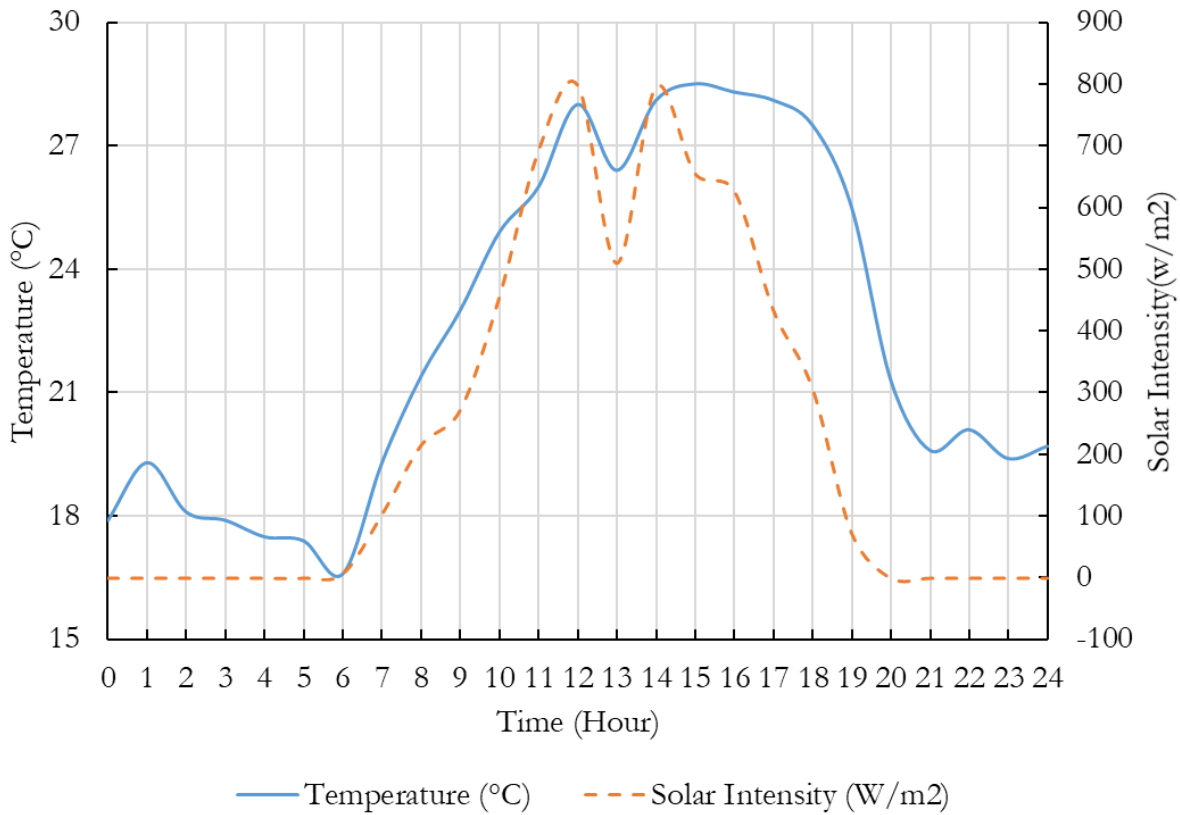

Figure 11. Local weather data on August 1st 2017 (NOAA 2017)

The time step size of the simulations is 30 seconds with 10 iterations in each time step. The simulation outputs were the temperatures on the concrete block surface. Both the models were validated with the experimental results. A mesh sensitivity was also conducted by comparing the results of the models with 100,000 and 10, 000 nodes. It was found that the difference in the concrete surface temperature results between the model with 100,000 and 10, 000 nodes were minor.

Based on the simulation (Figure 12), we observed results that indoor cooling case and outdoor case were consistent with experiment outcome. The simulation of indoor case shows high visibility for defects with size of 4 by 4, 6 by 4, 4 by 9 in depth of 1.5 and 2.5 for both heating and cooling phase. Furthermore, the defect size of 2 by 18 in 1.5 deep and the defect size of 4 by 9 and 6 by 6 in 3.5 deep were visible in the heating case. However, this maximal visibility was not observed in the experimental settings due to other indoor environmental noises. For the outdoor case, a low temperature contrast at 10 am was shown in the simulation (Figure 12. Bottom left) that the defect with sizes of 4 by 9 and 6 by 6 at depth of 1.5 and 2.5 inches were barely seen. The best outdoor visibility was shown in the 3 pm (Figure 12. Bottom middle) that defects with sizes of 4x 4, 4x 9, 6x 6, and 2x 18 at depth of 1.5 and 2.5 inches were seen. The result at 8 pm did not show visibility for any mimicked defects.

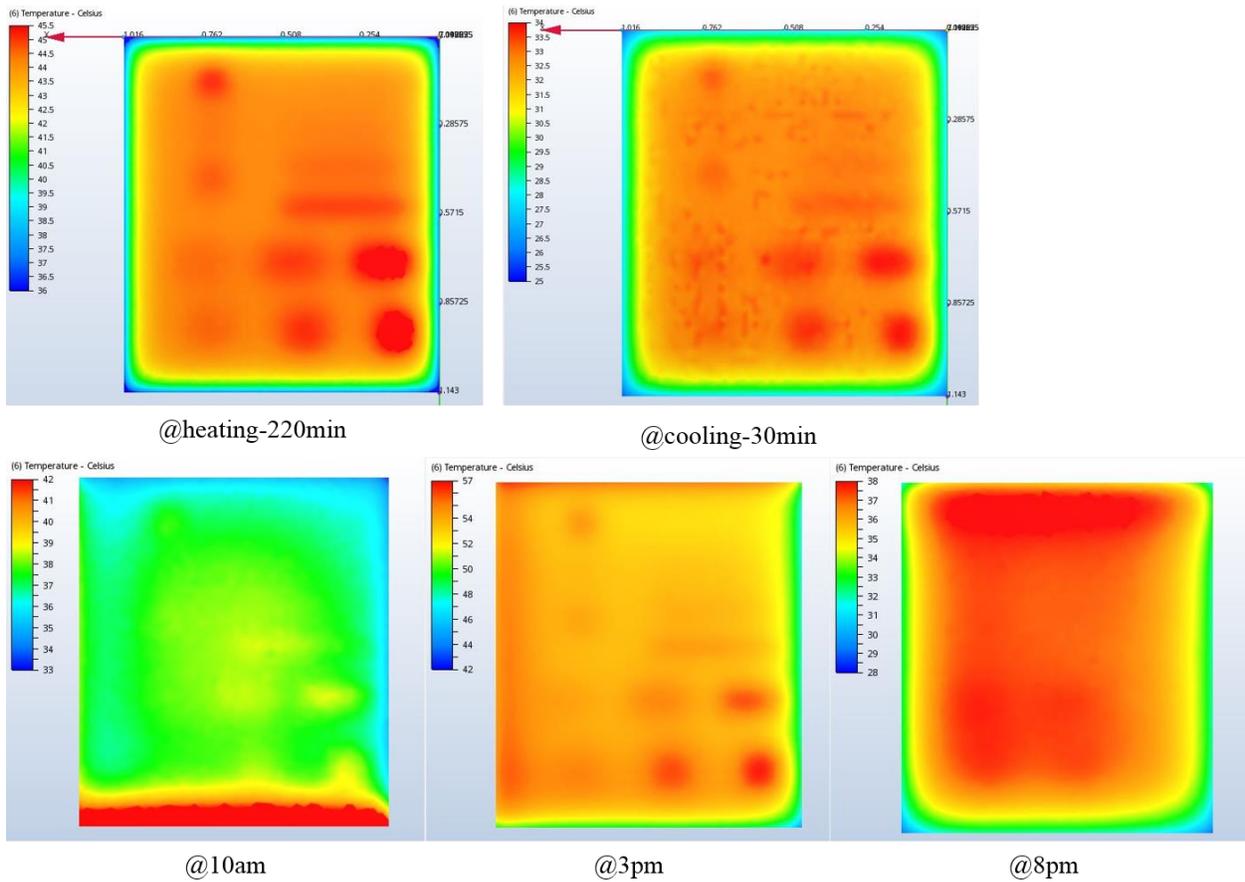

Figure 12. simulation results for (top) indoor condition and (bottom) outdoor condition (Each image has been scaled for the color palette to show best visibility)

# 7 Discussion and Conclusions

Since this paper provided an empirical procedure for thermal image content analysis in terms of delamination detection, the selection of LoG filter and determination of levels would be essential. We would recommend starting at small sigma value (such as 2 used in this paper) for the LoG filter. It would give the finer result at the early levels (Figure 7 L0 and L1) and provide some hints about surface textures at beginning then leading to more correlated pattern of delamination in the later levels. In terms of the stopping criterion for levels, it not only depends on the resolution of original image but also related to the temperature distribution across the raw image. With further decomposing the row images (Figure 6 (a), (b) and (c)) into level 5 (Figure 13), we could observe the patterns correlating to the concentrated heating sources. Thus, we stopped the further processing when this trending feature was occurred.

Considering both the experimental and the numerical simulation results, we concluded that the cooling stage with indoor setup, and afternoon time for outdoor environment are the best time of detecting delamination. By referring the simulation study done by Hiasa et al. (2017), the detectability varies for different sizes, buried depth, and thickness. For our case, the relatively small sizes of artificial defects may exhibit larger sensitivity to the time of a day and environmental influence. The indoor experiment performed better than the outdoor experiment due to primarily the stable room temperature. At outdoor environment, the temperature contrast was further reduced by clouding and wind conditions in terms of the relationship between sizes and depths. Compared to the conventional methods, the proposed procedure performed significantly better during the cooling the stage of indoor setup, and during the PM time under the natural environment. Compared to time-series methods for indoor environment, the proposed procedure showed its advantage by demonstrating similar detection performance, but only requires single image. It also found that the spatial feature of delamination was indeed embedded in the single thermal image which was masked by textural and non-uniform distributed noises and could be revealed by the proposed method. In outdoor environment, the time-series methods failed to generate meaningful result due to high environment-induced noise.

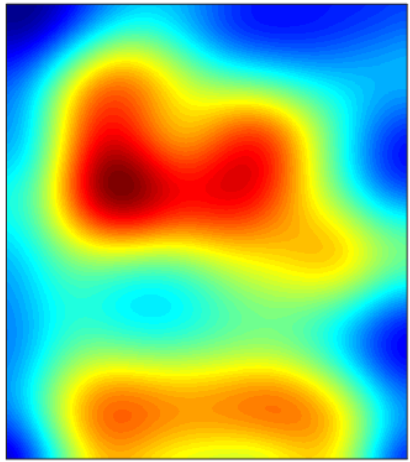 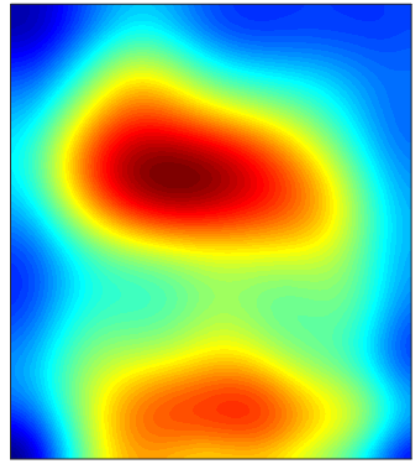

(a) (b)

Figure 13. Level 5 processed images revealing the heating source corresponding to Figure 6: (a) 218-minute at heating; (b) 30-minute at cooling

# 9 Literature Cited

Adelson, E. H., Anderson, C. H., Bergen, J. R., Burt, P. J., and Ogden, J. M. (1984). "Pyramid methods in image processing." *RCA engineer*, 29(6), 33-41.

Arndt, R. W. (2010). "Square pulse thermography in frequency domain as adaptation of pulsed phase thermography for qualitative and quantitative applications in cultural heritage and civil engineering." *Infrared Physics & Technology*, 53(4), 246-253.

Autodesk. 2017. Autodesk CFD [Computer software], Mill Valley, CA: Autodesk, 2017.

Dabous, S. A., Yaghi, S., Alkass, S., and Moselhi, O. (2017). "Concrete bridge deck condition assessment using IR Thermography and Ground Penetrating Radar technologies." *Automation in Construction*.

Dumoulin, J., and Averty, R. "Development of an infrared system coupled with a weather station for real time atmospheric corrections using GPU computing: Application to bridge monitoring." *Proc., Proc of 11th International Conference on Quantitative InfraRed Thermography, Naples Italy*.

Dumoulin, J., Crinière, A., and Averty, R. (2013). "The detection and thermal characterization of the inner structure of the 'Musmeci' bridge deck by infrared thermography monitoring." *Journal of Geophysics and Engineering*, 10(6), 064003.

Gholamalizadeh, E., and Kim, M.-H. (2014). "Three-dimensional CFD analysis for simulating the greenhouse effect in solar chimney power plants using a two-band radiation model." *Renewable energy*, 63, 498-506.

Glorot, X., Bordes, A., and Bengio, Y. "Deep sparse rectifier neural networks." *Proc., Proceedings of the Fourteenth International Conference on Artificial Intelligence and Statistics*, 315-323.

Heeger, D. J., and Bergen, J. R. "Pyramid-based texture analysis/synthesis." *Proc., Proceedings of the 22nd annual conference on Computer graphics and interactive techniques*, ACM, 229-238.

Hiasa, S. (2016). "Investigation of infrared thermography for subsurface damage detection of concrete structures."

Hiasa, S., Birgul, R., and Catbas, F. N. (2017). "A data processing methodology for infrared thermography images of concrete bridges." *Computers & Structures*, 190, 205-218.

Hiasa, S., Birgul, R., and Catbas, F. N. (2017). "Effect of defect size on subsurface defect detectability and defect depth estimation for concrete structures by infrared thermography." *Journal of Nondestructive Evaluation*, 36(3), 57.

Ibarra-Castanedo, C., and Maldague, X. (2004). "Pulsed phase thermography reviewed." *Quantitative Infrared Thermography Journal*, 1(1), 47-70.

Ibarra-Castanedo, C., and Maldague, X. "Review of pulsed phase thermography." *Proc., Thermosense: Thermal Infrared Applications XXXVII*, International Society for Optics and Photonics, 94850T.

Ibarra-Castanedo, C., Piau, J.-M., Guilbert, S., Avdelidis, N. P., Genest, M., Bendada, A., and Maldague, X. P. (2009). "Comparative study of active thermography techniques for the nondestructive evaluation of honeycomb structures." *Research in Nondestructive Evaluation*, 20(1), 1-31.

Ibarra-Castanedo, C., Sfarra, S., Klein, M., and Maldague, X. (2017). "Solar loading thermography: Time-lapsed thermographic survey and advanced thermographic signal processing for the inspection of civil engineering and cultural heritage structures." *Infrared Physics & Technology*, 82, 56-74.

Kong, H., Akakin, H. C., and Sarma, S. E. (2013). "A generalized Laplacian of Gaussian filter for blob detection and its applications." *IEEE transactions on cybernetics*, 43(6), 1719-1733.

Lindeberg, T. (1994). "Scale-space theory: A basic tool for analyzing structures at different scales." *Journal of applied statistics*, 21(1-2), 225-270.

Lindeberg, T. (1998). "Feature detection with automatic scale selection." *International journal of computer vision*, 30(2), 79-116.

Maierhofer, C., Arndt, R., Röllig, M., Rieck, C., Walther, A., Scheel, H., and Hillemeier, B. (2006). "Application of impulse-thermography for non-destructive assessment of concrete structures." *Cement and Concrete Composites*, 28(4), 393-401.

Maldague, X., Galmiche, F., and Ziadi, A. (2002). "Advances in pulsed phase thermography." *Infrared physics & technology*, 43(3), 175-181.

Martinopoulos, G., Missirlis, D., Tsilingiridis, G., Yakinthos, K., and Kyriakis, N. (2010). "CFD modeling of a polymer solar collector." *Renewable Energy*, 35(7), 1499-1508.

Milovanović, B., Banjad Pečur, I., and Štirmer, N. (2017). "The methodology for defect quantification in concrete using IR thermography." *Journal of Civil Engineering and Management*, 23(5), 573-582.

NOAA (2017). "Quality Controlled Datasets." The National Oceanic and Atmospheric Administration, National Climatic Data Center.

Omar, T., and Nehdi, M. L. (2017). "Remote sensing of concrete bridge decks using unmanned aerial vehicle infrared thermography." *Automation in Construction*, 83, 360-371.

Omar, T., Nehdi, M. L., and Zayed, T. (2017). "Rational Condition Assessment of RC Bridge Decks Subjected to Corrosion-Induced Delamination." *Journal of Materials in Civil Engineering*, 30(1), 04017259.

Selmi, M., Al-Khawaja, M. J., and Marafia, A. (2008). "Validation of CFD simulation for flat plate solar energy collector." *Renewable energy*, 33(3), 383-387.

Sultan, A. A., and Washer, G. (2017). "A pixel-by-pixel reliability analysis of infrared thermography (IRT) for the detection of subsurface delamination." *NDT & E International*, 92, 177-186.

Vollmer, M., and Möllmann, K.-P. (2010). *Infrared thermal imaging: fundamentals, research and applications*, John Wiley & Sons.

Zhang, M., Wu, T., and Bennett, K. M. (2015). "Small blob identification in medical images using regional features from optimum scale." *IEEE transactions on biomedical engineering*, 62(4), 1051-1062.

# 10 Appendix

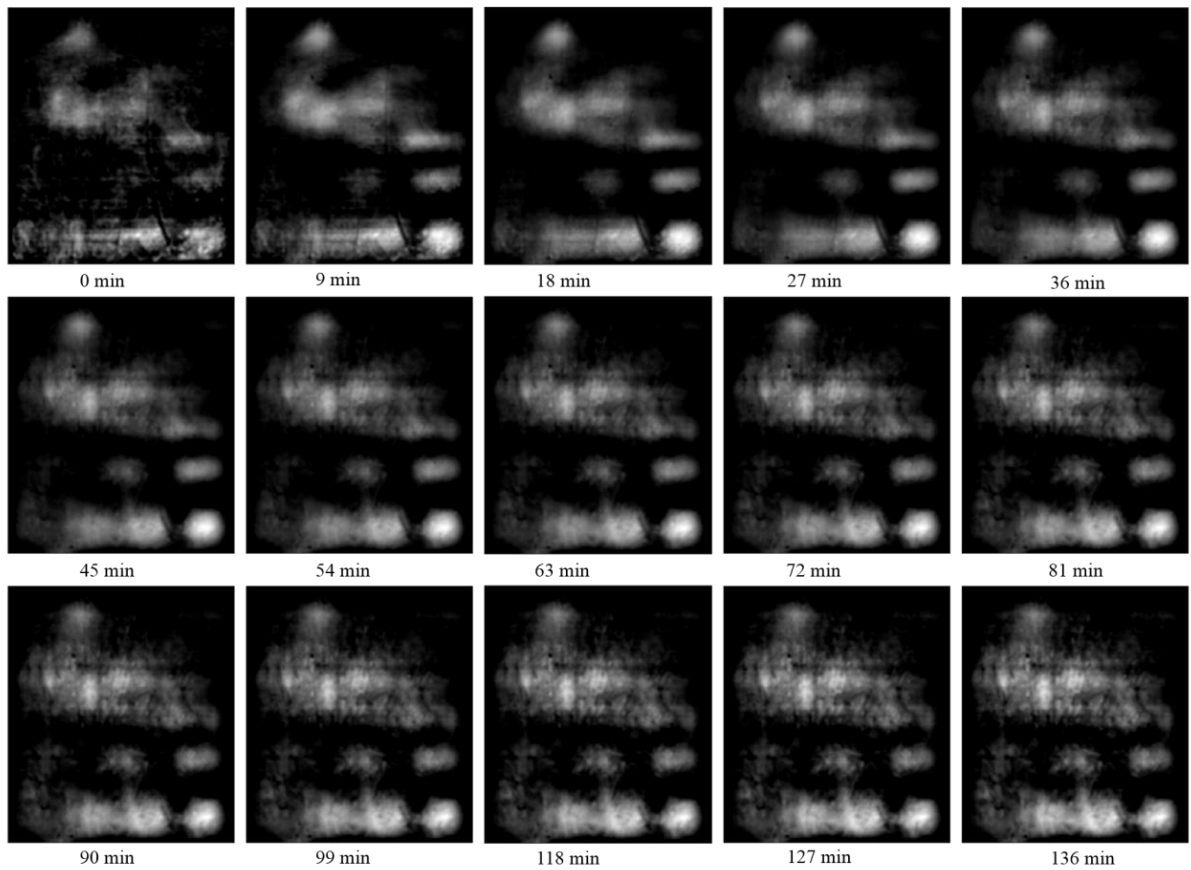

Figure I. processed images at different time windows for lamp case during the cooling phase

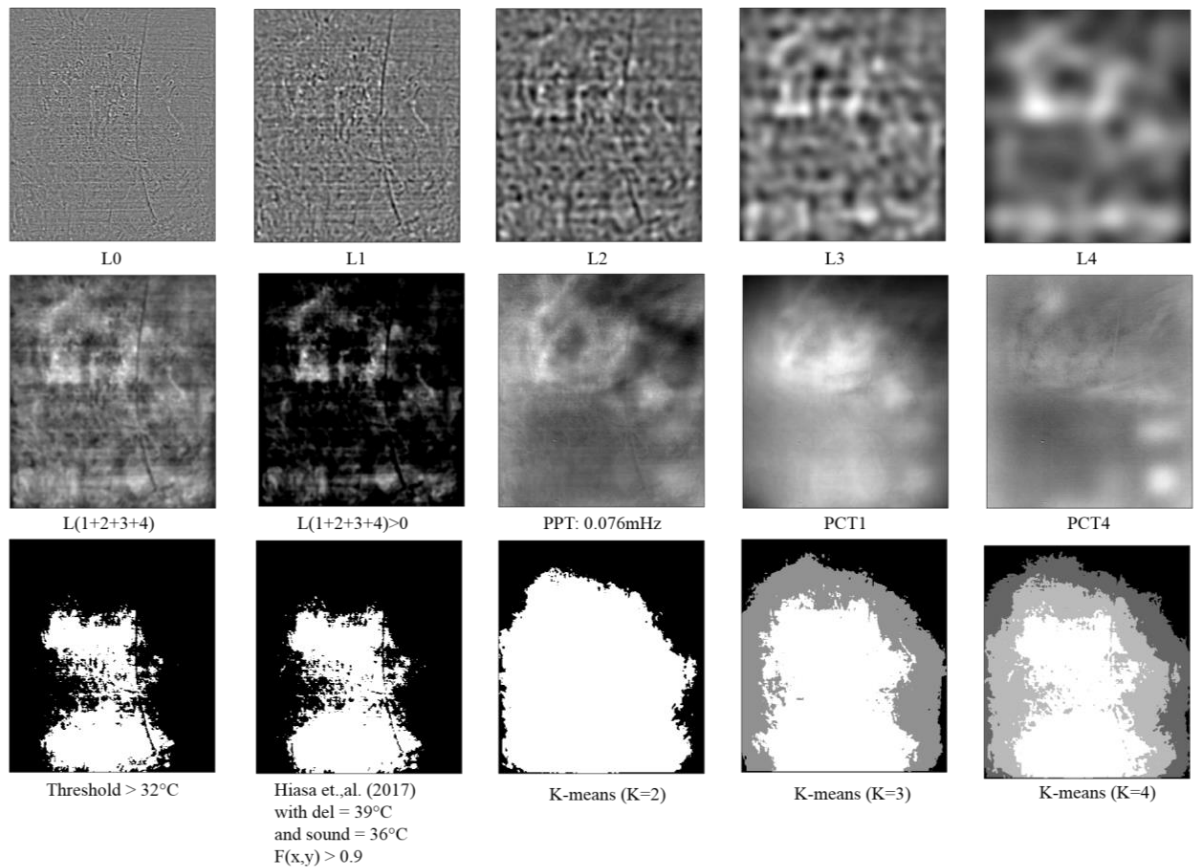

Figure II. Heating phase from lamp case (processed on Figure 6 (a))

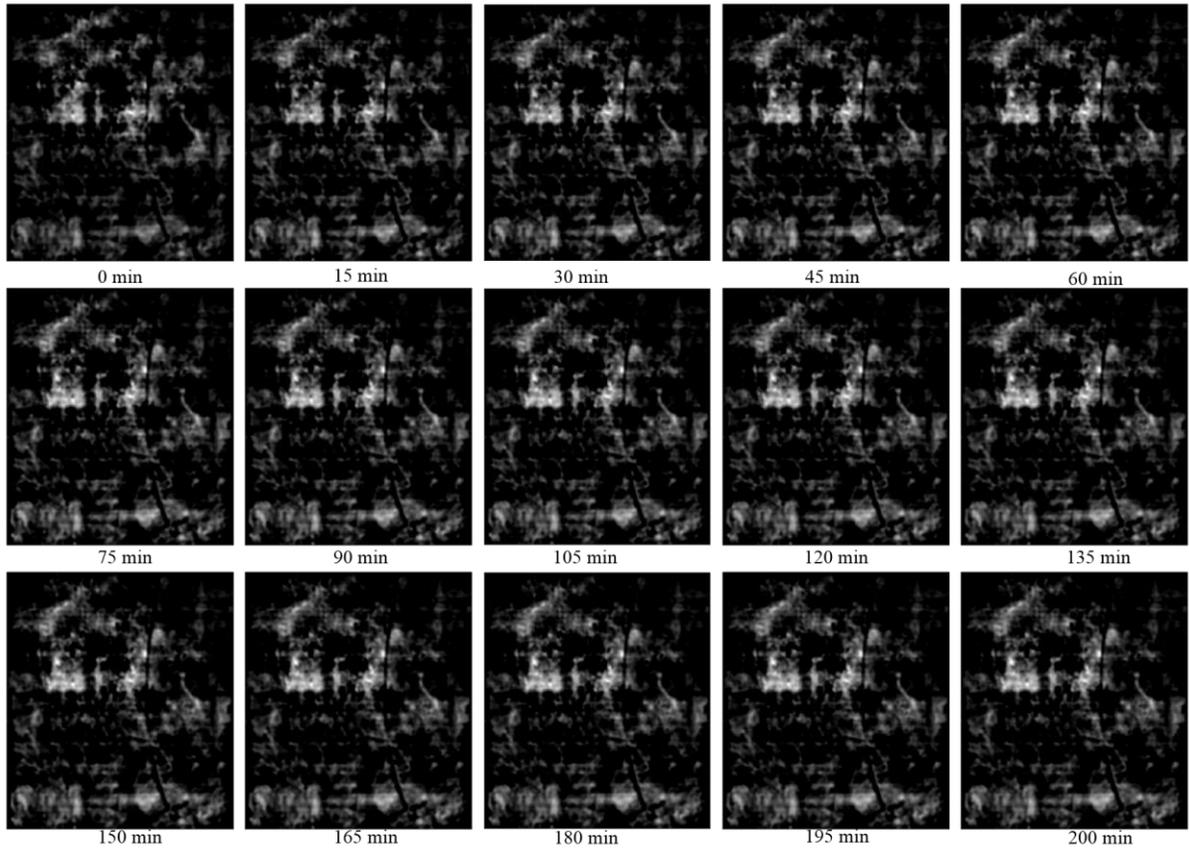

Figure III. processed images at different time windows for lamp case during the heating phase